# Channel Capacity of Coding System on Tsallis Entropy and $q$-Statistics

**Tatsuaki Tsuruyama**

Department of Pathology, Kyoto University, Graduate School of Medicine,

Yoshida-Konoe-cho, Sakyo-ku, Kyoto 606-8315, Japan; tsuruyam@kuhp.kyoto-u.ac.jp;

Tel.: +81-75-334-6252; Fax: +81-75-753-4493

**Abstract**

The field of information science has greatly developed, and applications in various fields have emerged. In this paper, we evaluated the coding system in the theory of Tsallis entropy for transmission of messages and aimed to formulate the channel capacity by maximization of the Tsallis entropy within a given condition of code length. As a result, we obtained a simple relational expression between code length and code appearance probability and, additionally, a generalized formula of the channel capacity on the basis of Tsallis entropy statistics. This theoretical framework may contribute to data processing techniques and other applications.

**Text**

Information theory has developed greatly in recent years, and has found broader applications in various research fields [1]. Shannon developed the information theory of entropy [2], and there have been extensive studies of entropy generalizations. On the other hand, the theory of Tsallis statistics originated in the 1980s, and the principle of entropy maximization was a means of expanding on the basic statistics theory. It was based on the fact that Boltzmann–Gibbs statistical mechanics could be reconstructed by the entropy maximization principle, a development beginning in 1957 with the work of Jaynes [3,4]; the framework has been developed primarily for extending statistical mechanics. Tsallis entropy was introduced in 1988 by Constantino Tsallis as a basis for generalizing the standard statistical mechanics on the basis of $q$-statistics [5]. It is possible to derive Tsallis distributions from the optimization of Tsallis entropy. For example, the $q$-Gaussian is one of the probability distributions that arise from the maximization of Tsallis entropy.

In this short paper, we aim to define the channel capacity of the coding system for transmission of messages on the basis of Tsallis entropy and aim to understand Tsallis entropy theory from the viewpoint of coding theory and seek a way to maximize the entropy corresponding to the signal event number. For the purpose, we introduce the coding symbols, the appearance probability of the symbols, and the message duration. Through this theory-based development, we reconsider the significance of Tsallis entropy in coding theory. Especially with respect to source coding, the objective is to determine the relational expression when applying the entropy maximization principle to establishing the relationship between code length and the appearance probability of the code used to calculate the channel capacity of a coding system. This theory regarding the relationship between code length and the appearance probability of the code was developed by Brillouin and the simple formula was obtained by maximizing information in the coding system, on the basis of information theory [6–8]. Several research progressions have been achieved regarding mutual information [7]; however, the relational expression between source code length and the appearance probability of the code remains to be improved in the theory.

## 2. Source Coding for Tsallis Entropy Formulation

Consider all the possible distinct messages that correspond to all the possible combinations of the code $A_j$, whose code length is $\tau_j$. For instance, a message is described using $n$ types of code symbols, $S_j$ ($1 \le j \le n$) as follows:

$$A_1\ A_3\ A_2\ A_3\ A_1\ A_4\ A_3\ A_5\ A_3\ A_3\ A_3 \qquad (1)$$

Our aim is to identify the way of coding in which the total information within a given duration can be maximized. The messages, which consist of symbols $A_j$ with numbers $N_j$ ($1 \leq j \leq n$), will correspond to all the possible combinations of symbols $A_j$. Therefore, $N_1 = 2$, $N_2 = 1$, $N_3 = 6$, and $n = 5$ in the message (1). Here, we consider $\Psi$, the total number of such distinct messages, in the selection of $n$ symbols. We assume absolutely no restrictions, constraints, or correlations in using various symbols. We obtain information $I$ derived from the above messages consisting of $N_j$,

$$I = K \log \psi \ . \qquad (2)$$

Here, $K$ is an arbitrary constant. If we use entropy unit, we take $K = k_B$, Boltzmann's constant. On the other hand, in information science, $K$ is equivalent to $\log_2 e$. Shannon defines the channel capacity as follows [2]:

$$C = \max \lim_{\tau \to \infty} \frac{K \log \psi}{\tau} = \max \lim_{\tau \to \infty} \frac{I}{\tau} \ . \qquad (3)$$

Here, $\Psi$ is signified as a function of $\tau$, a message of total duration. The unit of channel capacity is given by bits per second, if the total duration is measured in seconds. We define the total number of code symbols $N$ in a given message as:

$$N = \sum_{j=1}^{n} N_j \qquad (4)$$

For example, $N = 11$ in (1). Thus, $N$ is variable in different individual messages. Next, $p_j$ is the appearance probability of the $S_j$ symbol in the messages consisting of a total of $N$ symbols. In the following summations over $j$:

$$p_j \triangleq \frac{N_j}{N} \qquad (5)$$

and

$$\sum_{j=1}^{n} p_j = 1 \qquad (6)$$

Using (5), we can rewrite (3) as follows:

$$C = \max \lim_{\tau \to \infty} \left( -KN \sum_{j=1}^{n} p_j \log p_j \, / \, \tau \right) = \max \lim_{\tau \to \infty} \left( KS \, / \, \tau \right) \ . \qquad (7)$$

Here, $S$ represents the Shannon entropy, $S$. In this study, we investigate the channel capacity when the entropy is given by Tsallis entropy. Here, we introduce the *q-duration* of the message, as follows:

$$\tau_q = N \sum_{j=1}^{n} \phi_j \tau_j \tag{8}$$

$\tau_j$ signifies the $j$th code length. Here, we used escort probability $\phi_j$:

$$\phi_j \triangleq \frac{p_j^q}{c_q}, \tag{9}$$

In actuality,

$$\sum_{j=1}^{n} \phi_j = \sum_{j=1}^{n} \frac{p_j^q}{c_q} = 1 \tag{10}$$

For simplification, we use general notation in Tsallis statistics:

$$c_q \triangleq \sum_{j=1}^{n} p_j^q \tag{11}$$

Tsallis entropy is given by [6] (http://www.tsakkus,cat.cbpf.br/TEMUCO.pdf.):

$$S_q = N \sum_{j=1}^{n} \frac{p_j - p_j^q}{q-1} \tag{12}$$

The theory of Tsallis statistics, based on the generalized form of entropy $S_q$ (q$\in$**R**), when $q \to 1$, recovers the Shannon entropy:

$$S = -N \sum_{j=1}^{n} p_j \log p_j . \tag{13}$$

We aimed to maximize Tsallis entropy (12) [8], $S_q$, instead of Shannon entropy. Then, we introduced a function $G$, using non-determined parameters $\beta$ and $\gamma$, in reference to (8), (9), and (10):

$$G\left(p_1, p_2, \cdots p_n; N\right) \triangleq S_q - \beta \sum_{j=1}^{n} \phi_j - \gamma N \sum_{j=1}^{n} \phi_j \tau_j \tag{14}$$

Then

$$\frac{\partial}{\partial p_j} G\left(p_1, p_2, \cdots p_n; N\right) = N \frac{1 - qp_j^{q-1}}{q-1} - \left(\beta + \gamma N \tau_j\right) \frac{qp_j^{q-1}\left(-p_j^q + c_q\right)}{c_q^2} \tag{15}$$

$$\frac{\partial}{\partial N}G\left(p_1, p_2, \cdots p_n; N\right) = \sum_{j=1}^{n}\frac{p_j - p_j^q}{q-1} - \sum_{j=1}^{n}\gamma\frac{p_j^q}{c_q}\tau_j \tag{16}$$

For calculation of (15), we used

$$\frac{\partial \phi_j}{\partial p_j} = \frac{q p_j^{q-1}\left(-p_j^q + c_q\right)}{c_q^2} \tag{17}$$

For maximization of $G\left(p_1, p_2, \cdots p_n; N\right)$, setting the right sides of (15) and (16) equal to zero, we have:

$$\left(\beta + \gamma N\tau_j\right)\frac{q p_j^{q-1}\left(-p_j^q + c_q\right)}{c_q^2} = N\frac{1 - q p_j^{q-1}}{q-1} \tag{18}$$

$$\gamma\frac{p_j^q}{c_q}\tau_j = \frac{p_j - p_j^q}{q-1} \tag{19}$$

and solving the above equations with respect to undetermined coefficients β and γ, we have:

$$\beta = \frac{N c_q p_j^{-q}\left(p_j^q\left(-p_j + p_j^q\right)q + \left(p_j + q p_j - 2 q p_j^q\right)c_q\right)}{q(q-1)\left(-p_j^q + c_q\right)} \tag{20}$$

and

$$\gamma_q \triangleq \frac{\gamma}{c_q} = -\frac{1 - p_j^{1-q}}{(q-1)\tau_j} \tag{21}$$

Rewriting (21), using the $q$-logarithm function,

$$\log_q x = \frac{1 - x^{1-q}}{q-1}, \tag{22}$$

and we obtain from (8), (21), (22):

$$-\log_q p_j = \gamma_q \tau_j \tag{23}$$

Equation (23) implies that most probable code symbols must be short, while the improbable code may be long. In fact, when $q$ approaches 1, (23) gives the logarithm according to Brillouin's work using another constant $\gamma'$:

$$-\log p_j = \gamma'\tau_j \tag{24}$$

Thus, the probability of symbol appearance can be described using the Tsallis duration in (22), which is similar to the Shannon entropy coding in (24) [9]. The above result is explicitly a natural extension of Brillouin's theory, regarding the relationship between coding and Shannon entropy, to the concept of Tsallis entropy.

## 3. Channel capacity and Tsallis entropy

As shown in (23), $\gamma_q$ is equivalent to the Tsallis average entropy production rate $\sigma_q$ during the transmission of the message:

$$\gamma_q = -\log_q p_j / \tau_j \triangleq \sigma_q \qquad (25)$$

Our definition now yields the $q$-channel capacity $C_q$, in reference to (3), (7), (8), and (21) as follows:

$$C_q \triangleq \lim_{\tau_q \to \infty} \frac{-KN \sum_{j=1}^{n} p_j^q \log_q p_j}{\tau_q} = \lim_{\tau_q \to \infty} \frac{KN\sigma_q \sum_{j=1}^{n} p_j^q \tau_j}{\tau_q} = K_q \sigma_q \qquad (26)$$

with

$$K_q \triangleq K c_q \qquad (27)$$

Here, $K$ is an arbitrary constant. Therefore, the channel capacity has a dimension identical to the entropy production rate and is equivalent to Tsallis average entropy production rate. Thus, the above result is explicitly a natural extension of channel capacity on Tsallis entropy.

## Conclusions

In this short article, we achieved an important formulation between code length and appearance probability in Equation (26). In a similar way to how Shannon's entropy was extended to Tsallis entropy, the source-coding theory based on the former entropy by Brillouin [9, 10] was theoretically generalized for the theory based on Tsallis entropy [5, 7, 11]. On the other hand, it remains to be discussed how one should interpret $q$-duration in Equations (7) and (19). The duration of a given message event is generally shorter than the actual message duration; however, when the appearance probability distribution obeys a $q$-Gaussian, the $q$-duration is an indicator available for use in the analysis of the coding background, in place of the actual duration. From this perspective, we will further investigate the definition of the $q$-duration and the interpretation of the limitation of our calculation in future.

My theoretical attempt to generalize source coding can be applied to data management within the areas of data communications, processing, and conversion, particularly in the development of imaging applications. Further investigation is needed with regards to the tractability of Tsallis entropy and $q$-statistics in evaluating actual experimental data. We actually applied the medical imaging technique, following previous reports [12,13]; in future work, we look to investigate systems in which entropy is measurable and/or definable.

**Acknowledgments:** This work was supported by a Grant-in-Aid from the Ministry of Education, Culture, Sports, Science, and Technology of Japan (Synergy of Fluctuation and Structure: Quest for Universal Laws in Non-Equilibrium Systems, P2013-201 Grant-in-Aid for Scientific Research on Innovative Areas, MEXT, Japan). We thank Prof. Kenichi Yoshikawa of Doshisha University, for his advices.

**Conflicts of Interest:** The author declare no conflict of interest.